\documentclass[conference]{IEEEtran}
\IEEEoverridecommandlockouts

\usepackage{cite}
\usepackage{amsmath,amssymb,amsfonts, txfonts}
\usepackage{algorithmic}
\usepackage{graphicx}
\usepackage{textcomp}
\usepackage{xcolor}
\usepackage{braket}
\usepackage{booktabs}
\usepackage{multirow}
\usepackage{pifont}
\usepackage{tablefootnote}
\usepackage{caption, subcaption}
\usepackage[export]{adjustbox}
\usepackage{listing}
\usepackage{tcolorbox}
\usepackage{graphicx}
\usepackage{tikz}

\def\BibTeX{{\rm B\kern-.05em{\sc i\kern-.025em b}\kern-.08em
    T\kern-.1667em\lower.7ex\hbox{E}\kern-.125emX}}

\begin{document}
\title{Quantum error correction for kids}

\author{\IEEEauthorblockN{Richard A. Wolf}
\IEEEauthorblockA{\textit{Irish Center for High-End Computing}}}

\maketitle

\begin{abstract}
    No one should wait until college to get acquainted with core concepts of quantum information. Given the human bias of favouring the familiar over the unknown, early exposure to concepts of quantum information helps learners build stronger appetence for the field, as well as allowing them to develop an intuitive approach to it. In this work, I present an intuitive gamified approach to one of the core concepts in quantum error correction: the stabiliser formalism.
\end{abstract}


\section*{Context}

Early exposure to scientific concepts is considered to impact positively future engagement of the audience with the subject matter. Given the recognised urgency of educating a wider audience to quantum technologies and the pivotal role quantum error correction (QEC) plays within quantum computing, targeting early outreach towards this branch of the field is key.

With quantum training having been recognised as a priority by authorities \cite{kaur2022defining} and quantum technologies now having their own European competence framework \cite{greinert2024european, greinert2023towards}, a number of creative approaches to teaching and outreach has seen the day. On top of the traditional approach to curricula building, a number of alternatives have been suggested such as thought experiments \cite{thoughtXP}, behaviourism-inspired processes \cite{wolf2023brief} and games. This gamification \cite{seskirquantum} has been increasingly explored through video games \cite{chiofalo2023quantum, cheung2016quantum, kultima2021quantum}, interactive story-telling \cite{skult2021marriage}, quantum games \cite{chiofalo2022games} and a variety of other playful approaches \cite{seskirquantum}.

In terms of age range, however, most initiatives target the upper range of K12, leaving audiences below the age of 12 under-served. Yet exposing early K12 learners to general STEM \cite{dejarnette2012america, article} or even physics \cite{lazzeroni2021teaching} has been done successfully. Some initiatives exist to familiarise these younger audiences with quantum-related topics, but they remain limited \cite{franklin2020exploring}. 

QEC \cite{shor1995scheme} is considered to be one of the major bottlenecks now faced by quantum computing \cite{bluvstein2024logical}. The process of turning physical information into logical information to shield it from noise and potential errors isn't specific to quantum computing, however some specificities of quantum systems make it more challenging. Barriers such as the \textit{no cloning theorem} or the collapse of the wave function due to measurement make it impossible for some of the primitives of classical error correction to hold. Moreover, even if continuous quantum errors are digitised, they are still of two kinds -bit flips and phase flips-, compared with classical errors which only affect bit flips. 

Though not the most popular field of quantum computing for newcomers, QEC harbours core concepts which can in some cases be presented in a straightforward way. Simple families of codes such as \textit{stabiliser codes} -to which the popular \textit{surface code} \cite{Dennis_2002, kitaev2003fault} belong- relies on, among others, the concept of parity checks. This concept which comes from classical error correction \cite{gallager1962low}, is central in a number of codes, both classical and quantum, and because of how intuitive it can be constitutes a fair candidate for early learning.   

In this paper, I present a game which helps players build an intuitive grasp of the concept of parity checks used in the stabiliser formalism. A main game is presented with two extensions. The initial game mimics a situation of classical error correction while the last one reflects a quantum one. Each game can either be played stand-alone or in continuation from the other ones, complexifying the game play until reaching the quantum setup.

\subsection*{How to use}
This paper is meant as a resources for people working in quantum education wishing to tackle the topic of QEC with young audiences. It provides the rules and mechanics of a game meant to convey intuitively basic element of classical and quantum error correction. I also provide a non-exhaustive list of \textit{focus points} from which educators can pick and choose the aspects they are most interested in conveying in their game sessions.
Note that this paper is not meant as an introduction or presentation of QEC to the reader, it assumes the reader is familiar with the aspects of QEC being modelled in the game and wishes to convey the intuitions behind it to a young audience.
\section*{Game overview}
In this game of communication, Messengers try to deliver information to a Receiver while avoiding being caught by the Noise.

\subsection*{Teams and scoring}
The game will see 2 teams compete against each other: the Communication and the Noise teams. Players of the Communication team are divided into two groups: the Messengers and the Receiver. The way the Communication team can score a point is by having the Messengers successfully deliver their message to the Receiver. To do so, Messengers must cross the space (hereafter called the Channel\footnote{Channel: in reference to a \textit{communication channel}}) while avoiding the Noise. The Noise team scores a point by preventing the Communication team to score.

Note that the winner at the end of all rounds will be either Communication or Noise based on their total scores across all rounds. Players don't win individually, the team of either Communication or Noise does.

The game goes on for as many rounds as there are players, so that each player gets to be the Receiver once and the Noise at least once -depending on the number of Noise roles available-. Players should swap teams so they get to experience different roles in the game.

\subsubsection*{Teams}
The minimal number of players for the Communication team is 3 Messengers + 1 Receiver = 4. The minimal number of players for the Noise team is 1. Below is a table with suggested team split for growing numbers (up to 10 players):

\begin{table}[htbp]
    \centering
    \begin{tabular}{lll}
        \toprule
        \textbf{Communication} & \textbf{Noise} & \textbf{Total} \\
        \midrule
       4 (3 + 1)    & 1     & 5 \\
       5 (4 + 1)    & 1     & 6 \\
       6 (5 + 1)    & 1     & 7 \\
       6 (5 + 1)    & 2     & 8 \\
       7 (6 + 1)    & 2     & 9 \\
       8 (7 + 1)    & 2     & 10 \\
       8 (7 + 1)    & 3     & 11 \\
       9 (8 + 1)    & 3     & 10 \\
        \bottomrule
    \end{tabular}
     \caption{Suggested player split}
\end{table}

\subsection*{General setup}
While the game can be adapted to suit various group sizes and age ranges, I detail the core setup for the target group size and age, leaving the readers (and future players!) room for adaptations.

\begin{itemize}
    \item \textbf{Age range}: 7+
    \item \textbf{Number of players}: 5+
    \item \textbf{Materials} \begin{itemize}
        \item Two non-translucent bags.
        \item Marbles of 2 colours in sufficient numbers (if there are $n$ players, there should be $2(n-2)$ marbles of each colour).
        \item One or more dice (depending on variations).
    \end{itemize}
    \item \textbf{Environment}: large space for movement, either outdoors or indoors with alterations or a lot of free space. Especially in the version where they run, participants should be given enough room.
    \item \textbf{Game-play}: competitive, by team.
\end{itemize}

\begin{figure}[h]
    \centering
    \tikzset{every picture/.style={line width=0.75pt}} 

\begin{tikzpicture}[x=0.75pt,y=0.75pt,yscale=-1,xscale=1]

\draw  [color={rgb, 255:red, 74; green, 144; blue, 226 }  ,draw opacity=1 ] (186,99.86) -- (350.43,99.86) -- (350.43,199.86) -- (186,199.86) -- cycle ;
\draw  [color={rgb, 255:red, 126; green, 211; blue, 33 }  ,draw opacity=1 ][line width=2.25]  (143,184.64) .. controls (143,177.82) and (149.59,172.29) .. (157.71,172.29) .. controls (165.84,172.29) and (172.43,177.82) .. (172.43,184.64) .. controls (172.43,191.47) and (165.84,197) .. (157.71,197) .. controls (149.59,197) and (143,191.47) .. (143,184.64) -- cycle ; \draw  [color={rgb, 255:red, 126; green, 211; blue, 33 }  ,draw opacity=1 ][line width=2.25]  (151.24,180.44) .. controls (151.24,179.76) and (151.9,179.21) .. (152.71,179.21) .. controls (153.52,179.21) and (154.18,179.76) .. (154.18,180.44) .. controls (154.18,181.12) and (153.52,181.68) .. (152.71,181.68) .. controls (151.9,181.68) and (151.24,181.12) .. (151.24,180.44) -- cycle ; \draw  [color={rgb, 255:red, 126; green, 211; blue, 33 }  ,draw opacity=1 ][line width=2.25]  (161.25,180.44) .. controls (161.25,179.76) and (161.9,179.21) .. (162.72,179.21) .. controls (163.53,179.21) and (164.19,179.76) .. (164.19,180.44) .. controls (164.19,181.12) and (163.53,181.68) .. (162.72,181.68) .. controls (161.9,181.68) and (161.25,181.12) .. (161.25,180.44) -- cycle ; \draw  [color={rgb, 255:red, 126; green, 211; blue, 33 }  ,draw opacity=1 ][line width=2.25]  (150.36,189.59) .. controls (155.26,192.88) and (160.17,192.88) .. (165.07,189.59) ;
\draw  [color={rgb, 255:red, 126; green, 211; blue, 33 }  ,draw opacity=1 ][line width=2.25]  (143,149.64) .. controls (143,142.82) and (149.59,137.29) .. (157.71,137.29) .. controls (165.84,137.29) and (172.43,142.82) .. (172.43,149.64) .. controls (172.43,156.47) and (165.84,162) .. (157.71,162) .. controls (149.59,162) and (143,156.47) .. (143,149.64) -- cycle ; \draw  [color={rgb, 255:red, 126; green, 211; blue, 33 }  ,draw opacity=1 ][line width=2.25]  (151.24,145.44) .. controls (151.24,144.76) and (151.9,144.21) .. (152.71,144.21) .. controls (153.52,144.21) and (154.18,144.76) .. (154.18,145.44) .. controls (154.18,146.12) and (153.52,146.68) .. (152.71,146.68) .. controls (151.9,146.68) and (151.24,146.12) .. (151.24,145.44) -- cycle ; \draw  [color={rgb, 255:red, 126; green, 211; blue, 33 }  ,draw opacity=1 ][line width=2.25]  (161.25,145.44) .. controls (161.25,144.76) and (161.9,144.21) .. (162.72,144.21) .. controls (163.53,144.21) and (164.19,144.76) .. (164.19,145.44) .. controls (164.19,146.12) and (163.53,146.68) .. (162.72,146.68) .. controls (161.9,146.68) and (161.25,146.12) .. (161.25,145.44) -- cycle ; \draw  [color={rgb, 255:red, 126; green, 211; blue, 33 }  ,draw opacity=1 ][line width=2.25]  (150.36,154.59) .. controls (155.26,157.88) and (160.17,157.88) .. (165.07,154.59) ;
\draw  [color={rgb, 255:red, 126; green, 211; blue, 33 }  ,draw opacity=1 ][line width=2.25]  (143,115.64) .. controls (143,108.82) and (149.59,103.29) .. (157.71,103.29) .. controls (165.84,103.29) and (172.43,108.82) .. (172.43,115.64) .. controls (172.43,122.47) and (165.84,128) .. (157.71,128) .. controls (149.59,128) and (143,122.47) .. (143,115.64) -- cycle ; \draw  [color={rgb, 255:red, 126; green, 211; blue, 33 }  ,draw opacity=1 ][line width=2.25]  (151.24,111.44) .. controls (151.24,110.76) and (151.9,110.21) .. (152.71,110.21) .. controls (153.52,110.21) and (154.18,110.76) .. (154.18,111.44) .. controls (154.18,112.12) and (153.52,112.68) .. (152.71,112.68) .. controls (151.9,112.68) and (151.24,112.12) .. (151.24,111.44) -- cycle ; \draw  [color={rgb, 255:red, 126; green, 211; blue, 33 }  ,draw opacity=1 ][line width=2.25]  (161.25,111.44) .. controls (161.25,110.76) and (161.9,110.21) .. (162.72,110.21) .. controls (163.53,110.21) and (164.19,110.76) .. (164.19,111.44) .. controls (164.19,112.12) and (163.53,112.68) .. (162.72,112.68) .. controls (161.9,112.68) and (161.25,112.12) .. (161.25,111.44) -- cycle ; \draw  [color={rgb, 255:red, 126; green, 211; blue, 33 }  ,draw opacity=1 ][line width=2.25]  (150.36,120.59) .. controls (155.26,123.88) and (160.17,123.88) .. (165.07,120.59) ;
\draw  [color={rgb, 255:red, 208; green, 2; blue, 27 }  ,draw opacity=1 ][line width=2.25]  (252,148.64) .. controls (252,141.82) and (258.59,136.29) .. (266.71,136.29) .. controls (274.84,136.29) and (281.43,141.82) .. (281.43,148.64) .. controls (281.43,155.47) and (274.84,161) .. (266.71,161) .. controls (258.59,161) and (252,155.47) .. (252,148.64) -- cycle ; \draw  [color={rgb, 255:red, 208; green, 2; blue, 27 }  ,draw opacity=1 ][line width=2.25]  (260.24,144.44) .. controls (260.24,143.76) and (260.9,143.21) .. (261.71,143.21) .. controls (262.52,143.21) and (263.18,143.76) .. (263.18,144.44) .. controls (263.18,145.12) and (262.52,145.68) .. (261.71,145.68) .. controls (260.9,145.68) and (260.24,145.12) .. (260.24,144.44) -- cycle ; \draw  [color={rgb, 255:red, 208; green, 2; blue, 27 }  ,draw opacity=1 ][line width=2.25]  (270.25,144.44) .. controls (270.25,143.76) and (270.9,143.21) .. (271.72,143.21) .. controls (272.53,143.21) and (273.19,143.76) .. (273.19,144.44) .. controls (273.19,145.12) and (272.53,145.68) .. (271.72,145.68) .. controls (270.9,145.68) and (270.25,145.12) .. (270.25,144.44) -- cycle ; \draw  [color={rgb, 255:red, 208; green, 2; blue, 27 }  ,draw opacity=1 ][line width=2.25]  (259.36,153.59) .. controls (264.26,156.88) and (269.17,156.88) .. (274.07,153.59) ;
\draw  [color={rgb, 255:red, 144; green, 19; blue, 254 }  ,draw opacity=1 ][line width=2.25]  (363,147.64) .. controls (363,140.82) and (369.59,135.29) .. (377.71,135.29) .. controls (385.84,135.29) and (392.43,140.82) .. (392.43,147.64) .. controls (392.43,154.47) and (385.84,160) .. (377.71,160) .. controls (369.59,160) and (363,154.47) .. (363,147.64) -- cycle ; \draw  [color={rgb, 255:red, 144; green, 19; blue, 254 }  ,draw opacity=1 ][line width=2.25]  (371.24,143.44) .. controls (371.24,142.76) and (371.9,142.21) .. (372.71,142.21) .. controls (373.52,142.21) and (374.18,142.76) .. (374.18,143.44) .. controls (374.18,144.12) and (373.52,144.68) .. (372.71,144.68) .. controls (371.9,144.68) and (371.24,144.12) .. (371.24,143.44) -- cycle ; \draw  [color={rgb, 255:red, 144; green, 19; blue, 254 }  ,draw opacity=1 ][line width=2.25]  (381.25,143.44) .. controls (381.25,142.76) and (381.9,142.21) .. (382.72,142.21) .. controls (383.53,142.21) and (384.19,142.76) .. (384.19,143.44) .. controls (384.19,144.12) and (383.53,144.68) .. (382.72,144.68) .. controls (381.9,144.68) and (381.25,144.12) .. (381.25,143.44) -- cycle ; \draw  [color={rgb, 255:red, 144; green, 19; blue, 254 }  ,draw opacity=1 ][line width=2.25]  (370.36,152.59) .. controls (375.26,155.88) and (380.17,155.88) .. (385.07,152.59) ;

\end{tikzpicture}
        \caption{Player setup\\ 
        \footnotesize{The Messengers (green) will have to cross the channel (rectangle) and reach the Receiver (purple). All this while avoiding to get caught by the Noise (red).}}
    \label{fig:initialSetup}
\end{figure}
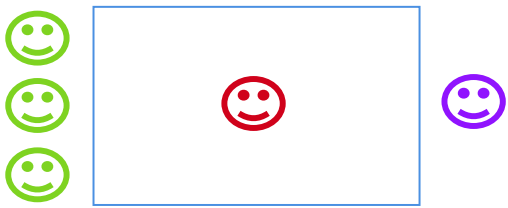
\section*{Repetition code}

In this basic version of the game, the Communication team carries information across the channel, avoiding being caught by the Noise, and delivers it to the Receiver. The Receiver then decodes the message and must guess what message was sent. 

\subsection*{Round setup}
Before the game starts, players prepare and are then asked to stand in position.
\begin{itemize}
    \item The Receiver stands on one side of the Channel, holding an empty bag.
    \item The Noise stands in the middle of the Channel with a die in their pocket.
    \item The Messengers stand on the side of the Channel that is across from the Receiver with a bag containing marbles of each of the 2 colours.
    \item Messengers agree on a colour for the round. Once agreed, each of them takes a marble of the round's colour with them.
    \item All players agree on a number called the \textbf{Noise Value}. When the Noise throws the die, any value equal or greater than the Noise Value means the Noise gets to act. Any value below means they don't.
\end{itemize}

\subsection*{Game phases}

\begin{figure}[h]
    \centering
    \tikzset{every picture/.style={line width=0.75pt}} 

\begin{tikzpicture}[x=0.75pt,y=0.75pt,yscale=-1,xscale=1]

\draw  [color={rgb, 255:red, 74; green, 144; blue, 226 }  ,draw opacity=1 ] (206,119.86) -- (370.43,119.86) -- (370.43,219.86) -- (206,219.86) -- cycle ;
\draw  [color={rgb, 255:red, 126; green, 211; blue, 33 }  ,draw opacity=1 ][line width=2.25]  (295,195.64) .. controls (295,188.82) and (301.59,183.29) .. (309.71,183.29) .. controls (317.84,183.29) and (324.43,188.82) .. (324.43,195.64) .. controls (324.43,202.47) and (317.84,208) .. (309.71,208) .. controls (301.59,208) and (295,202.47) .. (295,195.64) -- cycle ; \draw  [color={rgb, 255:red, 126; green, 211; blue, 33 }  ,draw opacity=1 ][line width=2.25]  (303.24,191.44) .. controls (303.24,190.76) and (303.9,190.21) .. (304.71,190.21) .. controls (305.52,190.21) and (306.18,190.76) .. (306.18,191.44) .. controls (306.18,192.12) and (305.52,192.68) .. (304.71,192.68) .. controls (303.9,192.68) and (303.24,192.12) .. (303.24,191.44) -- cycle ; \draw  [color={rgb, 255:red, 126; green, 211; blue, 33 }  ,draw opacity=1 ][line width=2.25]  (313.25,191.44) .. controls (313.25,190.76) and (313.9,190.21) .. (314.72,190.21) .. controls (315.53,190.21) and (316.19,190.76) .. (316.19,191.44) .. controls (316.19,192.12) and (315.53,192.68) .. (314.72,192.68) .. controls (313.9,192.68) and (313.25,192.12) .. (313.25,191.44) -- cycle ; \draw  [color={rgb, 255:red, 126; green, 211; blue, 33 }  ,draw opacity=1 ][line width=2.25]  (302.36,200.59) .. controls (307.26,203.88) and (312.17,203.88) .. (317.07,200.59) ;
\draw  [color={rgb, 255:red, 126; green, 211; blue, 33 }  ,draw opacity=1 ][line width=2.25]  (318,156.64) .. controls (318,149.82) and (324.59,144.29) .. (332.71,144.29) .. controls (340.84,144.29) and (347.43,149.82) .. (347.43,156.64) .. controls (347.43,163.47) and (340.84,169) .. (332.71,169) .. controls (324.59,169) and (318,163.47) .. (318,156.64) -- cycle ; \draw  [color={rgb, 255:red, 126; green, 211; blue, 33 }  ,draw opacity=1 ][line width=2.25]  (326.24,152.44) .. controls (326.24,151.76) and (326.9,151.21) .. (327.71,151.21) .. controls (328.52,151.21) and (329.18,151.76) .. (329.18,152.44) .. controls (329.18,153.12) and (328.52,153.68) .. (327.71,153.68) .. controls (326.9,153.68) and (326.24,153.12) .. (326.24,152.44) -- cycle ; \draw  [color={rgb, 255:red, 126; green, 211; blue, 33 }  ,draw opacity=1 ][line width=2.25]  (336.25,152.44) .. controls (336.25,151.76) and (336.9,151.21) .. (337.72,151.21) .. controls (338.53,151.21) and (339.19,151.76) .. (339.19,152.44) .. controls (339.19,153.12) and (338.53,153.68) .. (337.72,153.68) .. controls (336.9,153.68) and (336.25,153.12) .. (336.25,152.44) -- cycle ; \draw  [color={rgb, 255:red, 126; green, 211; blue, 33 }  ,draw opacity=1 ][line width=2.25]  (325.36,161.59) .. controls (330.26,164.88) and (335.17,164.88) .. (340.07,161.59) ;
\draw  [color={rgb, 255:red, 126; green, 211; blue, 33 }  ,draw opacity=1 ][line width=2.25]  (220,147.64) .. controls (220,140.82) and (226.59,135.29) .. (234.71,135.29) .. controls (242.84,135.29) and (249.43,140.82) .. (249.43,147.64) .. controls (249.43,154.47) and (242.84,160) .. (234.71,160) .. controls (226.59,160) and (220,154.47) .. (220,147.64) -- cycle ; \draw  [color={rgb, 255:red, 126; green, 211; blue, 33 }  ,draw opacity=1 ][line width=2.25]  (228.24,143.44) .. controls (228.24,142.76) and (228.9,142.21) .. (229.71,142.21) .. controls (230.52,142.21) and (231.18,142.76) .. (231.18,143.44) .. controls (231.18,144.12) and (230.52,144.68) .. (229.71,144.68) .. controls (228.9,144.68) and (228.24,144.12) .. (228.24,143.44) -- cycle ; \draw  [color={rgb, 255:red, 126; green, 211; blue, 33 }  ,draw opacity=1 ][line width=2.25]  (238.25,143.44) .. controls (238.25,142.76) and (238.9,142.21) .. (239.72,142.21) .. controls (240.53,142.21) and (241.19,142.76) .. (241.19,143.44) .. controls (241.19,144.12) and (240.53,144.68) .. (239.72,144.68) .. controls (238.9,144.68) and (238.25,144.12) .. (238.25,143.44) -- cycle ; \draw  [color={rgb, 255:red, 126; green, 211; blue, 33 }  ,draw opacity=1 ][line width=2.25]  (227.36,152.59) .. controls (232.26,155.88) and (237.17,155.88) .. (242.07,152.59) ;
\draw  [color={rgb, 255:red, 208; green, 2; blue, 27 }  ,draw opacity=1 ][line width=2.25]  (275,177.64) .. controls (275,170.82) and (281.59,165.29) .. (289.71,165.29) .. controls (297.84,165.29) and (304.43,170.82) .. (304.43,177.64) .. controls (304.43,184.47) and (297.84,190) .. (289.71,190) .. controls (281.59,190) and (275,184.47) .. (275,177.64) -- cycle ; \draw  [color={rgb, 255:red, 208; green, 2; blue, 27 }  ,draw opacity=1 ][line width=2.25]  (283.24,173.44) .. controls (283.24,172.76) and (283.9,172.21) .. (284.71,172.21) .. controls (285.52,172.21) and (286.18,172.76) .. (286.18,173.44) .. controls (286.18,174.12) and (285.52,174.68) .. (284.71,174.68) .. controls (283.9,174.68) and (283.24,174.12) .. (283.24,173.44) -- cycle ; \draw  [color={rgb, 255:red, 208; green, 2; blue, 27 }  ,draw opacity=1 ][line width=2.25]  (293.25,173.44) .. controls (293.25,172.76) and (293.9,172.21) .. (294.72,172.21) .. controls (295.53,172.21) and (296.19,172.76) .. (296.19,173.44) .. controls (296.19,174.12) and (295.53,174.68) .. (294.72,174.68) .. controls (293.9,174.68) and (293.25,174.12) .. (293.25,173.44) -- cycle ; \draw  [color={rgb, 255:red, 208; green, 2; blue, 27 }  ,draw opacity=1 ][line width=2.25]  (282.36,182.59) .. controls (287.26,185.88) and (292.17,185.88) .. (297.07,182.59) ;
\draw  [color={rgb, 255:red, 144; green, 19; blue, 254 }  ,draw opacity=1 ][line width=2.25]  (383,167.64) .. controls (383,160.82) and (389.59,155.29) .. (397.71,155.29) .. controls (405.84,155.29) and (412.43,160.82) .. (412.43,167.64) .. controls (412.43,174.47) and (405.84,180) .. (397.71,180) .. controls (389.59,180) and (383,174.47) .. (383,167.64) -- cycle ; \draw  [color={rgb, 255:red, 144; green, 19; blue, 254 }  ,draw opacity=1 ][line width=2.25]  (391.24,163.44) .. controls (391.24,162.76) and (391.9,162.21) .. (392.71,162.21) .. controls (393.52,162.21) and (394.18,162.76) .. (394.18,163.44) .. controls (394.18,164.12) and (393.52,164.68) .. (392.71,164.68) .. controls (391.9,164.68) and (391.24,164.12) .. (391.24,163.44) -- cycle ; \draw  [color={rgb, 255:red, 144; green, 19; blue, 254 }  ,draw opacity=1 ][line width=2.25]  (401.25,163.44) .. controls (401.25,162.76) and (401.9,162.21) .. (402.72,162.21) .. controls (403.53,162.21) and (404.19,162.76) .. (404.19,163.44) .. controls (404.19,164.12) and (403.53,164.68) .. (402.72,164.68) .. controls (401.9,164.68) and (401.25,164.12) .. (401.25,163.44) -- cycle ; \draw  [color={rgb, 255:red, 144; green, 19; blue, 254 }  ,draw opacity=1 ][line width=2.25]  (390.36,172.59) .. controls (395.26,175.88) and (400.17,175.88) .. (405.07,172.59) ;
\draw   (232.43,105.11) -- (299.97,105.11) -- (299.97,101.86) -- (345,108.36) -- (299.97,114.86) -- (299.97,111.61) -- (232.43,111.61) -- (235.68,108.36) -- cycle ;

\end{tikzpicture}
        \caption{Crossing the channel phase\\ 
    \footnotesize{Messengers (green) start crossing the channel trying to avoid the Noise (red). Here one of the Messenger was caught by the noise}}
    \label{fig:crossingChannel}
\end{figure}
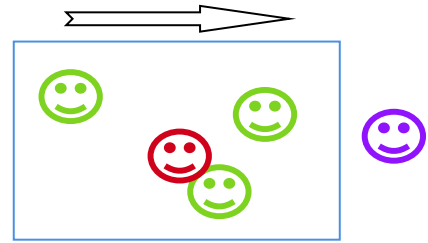

\subsubsection*{Crossing the Channel}
A member of the Noise gives the GO signal. As soon as the signal is given, Messengers can start crossing the Channel. The goal of the Noise is to catch as many Messengers as they can before they reach the Receiver. The goal of the Messengers is to reach the Receiver before getting caught by the Noise.
If a Messenger gets caught by the Noise, they must freeze in place and wait for the crossing to be over.
If a Messenger manages to reach the Receiver, they put their marble in the Receiver's bag. It's crucial at this stage that the Receiver does \textbf{not} look at the marble or inside the bag.
Once all Messengers have either reached the Receiver or are have been frozen in place by the Noise, this phase is over.

\subsubsection*{Noise acts}
In the case where the Noise failed to catch any Messenger, this phase is skipped. In the event that the Noise did catch some Messengers, their turn now begins. 
Noise collects the marble bag from the Messenger's side of the Channel and approaches each Messenger one by one. For each Messenger, the interaction goes as follows:
\begin{itemize}
    \item The Receiver should keep their eyes closed during each of the interactions. It is important at this stage that the Receiver does not see whether the Noise exchanges the Messenger's marbles or not.
    \item The Noise rolls the die. If the outcome is below the noise value, nothing happens. If the result is equal to or higher than the noise value, the Noise swaps the marble of the Messenger for a marble of the opposite colour.
\end{itemize}
Once the Noise has interacted with each Messenger once, all Messengers join the Receiver, put their marble in the bag, and wait until the Decoding phase.
Once all the Messengers have put their marbles into the Receiver's bag, this phase is over.

\subsubsection*{Decoding}

\begin{figure}[h]
    \centering
    \input{images/decoding_simple}
        \caption{Example decoding phase\\ 
    \footnotesize{The Receiver (purple) evaluates whether the message brought by the Messengers is \textit{black} or \textit{white}. On the left, 2 black marbles and a white one would suggest the original message is likely to have been black. Similarly on the right, the original message is likely to have been \textit{white}. }}
    \label{fig:decodingSimple}
\end{figure}

It is time for the Receiver to decode the message. The Receiver can now look into the bag and take the marbles out. They must guess which was the initial colour. If they guess correctly, the Communication team scores a point. If they guess incorrectly, the Noise team scores a point.
\section*{Parity checks}
This version can be either proposed as a stand-alone game or as a modification of the previous one. Here, the goal of the Receiver changes. Instead of figuring out whether the initial message was \textit{black} or \textit{white}, the Receiver's aim is to identify \textbf{\textit{where}} the error happened. if the Receiver guesses correctly where the error-s happened, the Communication team scores a point, if they don't the Noise does.
The game-play follows the general flow of the \textit{Repetition code} except for the following changes.

\begin{itemize}
    \item \textbf{Additional material}: red and green colour cards.
\end{itemize}

\subsection*{Noise phase}
In this variant, the Receiver must keep their eyes closed during the whole Noise phase.

\subsection*{Decoding phase}
When reaching the Receiver, Messengers do not put their marbles into the bag but simply wait.
During the Decoding phase, the Receiver will call Messengers two by two. They can call any pair combination any number of time they want. Now, instead of looking at the colours of their marbles, the Receiver is only allowed to ask whether their marbles are the same colour or whether they are not. Upon being asked to compare their marble colours, players look at each other's marble colour (without the Receiver seeing them). If their marbles are the same colour, one of them pulls out their green card and shows it to the Receiver. If their marbles are of different colours, one of them pulls the red card and show it to the Receiver. Based on this information, the Receiver is meant to identify which Messengers they think were affected by Noise.

\begin{figure}[h]
    \centering
    \input{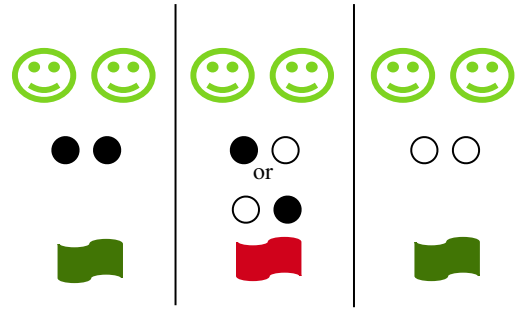}
        \caption{Correspondence between possible marble pairs and card colour selection.
    }
    \label{fig:decodingSimple}
\end{figure}

\begin{figure}[h]
    \centering
    \input{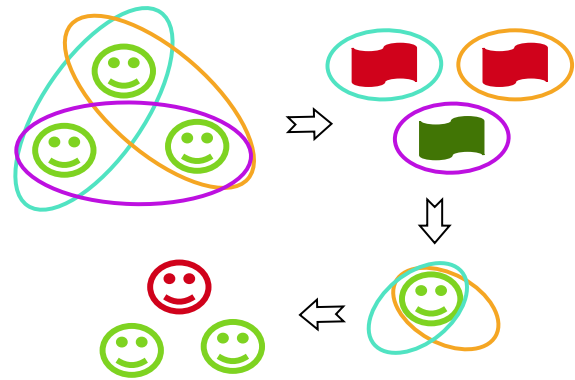}
        \caption{Example of the induction process for parity checks.\\
        \footnotesize{
        For 3 Messengers, the Receiver has 3 possible parities to check: orange, blue and pink. Each of these pair of players will show a card. Here both the orange and the blue checks show a red card while the pink one shows a green card. Knowing this, the Receiver knows the error has happened on the information of the Messenger which is part of both of these, but not of the pink one. The only Messenger which belongs to both the blue and orange parity checks but not to the pink one is the one on top. This is where the error happened. 
        }
        }
    \label{fig:parityCheck}
\end{figure}
\section*{Quantum codes}
Once again, this game can be played either as a stand-alone or an extension of the previous ones.

\begin{itemize}
    \item \textbf{Additional material}: black and white flat marbles, $2(n-2)$ of each colour.
\end{itemize}

This last version of the game is the one which actually mimics the protection of quantum information, as opposed to the previous versions which were classical error correction schemes. As such it is slightly more complex than the previous ones and players should be made familiar with the game mechanics through the first versions before moving on to this one.

The game unfolds in a similar way to the previous \textit{Parity checks} ones, with some modifications.

\subsection*{Noise phase}
Instead of only being able to exchange marble colours, the Noise can now either exchange either the colour or the shape of the marble, or both. The Noise first throws to see if they affect the colour of the marble. Here the same rule applies as before when checking the Noise Value. Then they throw the die a second time for the shape.

\subsection*{Decoding phase}
Now instead of having to figure out only which colours have been switched, the receiver also has to figure out which shapes have been. The actions allowed to figure that out are the same as in the \textit{Parity check} version except that the Receiver can now either ask whether players share the same shape or share the same colours.
\section*{Game modifications}

\subsection*{Simplifications}
If the Noise has difficulty catching Messengers, you can either change the ratio and add more Noise players to the team (though it is not recommended for less than 6 players) or increase the probability that the Noise can act if it catches a player by lowering the Noise Value or changing the die.

\subsection*{Complexifications}

\subsubsection*{Exploring different Messengers/Noise proportions}
The proportion of players being the Messengers or the Noise can be varied.

\subsubsection*{Exploration of noise levels}
Players can experiment by changing the Noise Value to different numbers and/or by using different dice.

\subsubsection*{Adversarial noise levels}
Instead of all players agreeing on a Noise Value, this choice could be left to the Noise alone. Instead of being publicly shared, it would also be possible for the Noise to keep that choice secret from the Receiver.

\subsubsection*{Measurement noise}
On top of the initial Noise phase which happens when Messengers cross the Channel, an additional phase for a different kind of Noise (measurement noise) can be accommodated. In this case, once all Messengers have put their marbles into the Receiver's bag, there is a last phase before the decoding one where the Noise can take swap some number of marbles for others from outside the bag. Whether they are swapped for the opposite colour or whether they are swapped for a random colour is left to the players to decide.

\subsubsection*{Limited resources}
In the \textit{Parity check} and \textit{Quantum code} versions, this modification encourages the Receiver to optimise information gain. This system requires the setting up of a points budget, and here asking one question to a pair would cost a negative point. In this scenario, the player is prompted to limit the number of checks being performed.

\subsubsection*{Stabiliser size}
With enough players, it is possible to allow the Receiver to experiment with different parity checks. For instance, instead of allowing only to ask questions to pairs of Messengers, the Receiver could be given the choice between asking a group of two or a group of four Messengers (or a group of six in larger settings). In this version, it will be important to emphasise to Messengers that they should count the \textbf{\textit{parity}} of their pieces of information and not whether or not they all agree. This confusion can arise when going from two to four players since a parity check for a two-player group is effectively the same as asking whether or not all elements are the same. Which is not the case for larger groups.

\begin{figure}[h]
    \centering
    \input{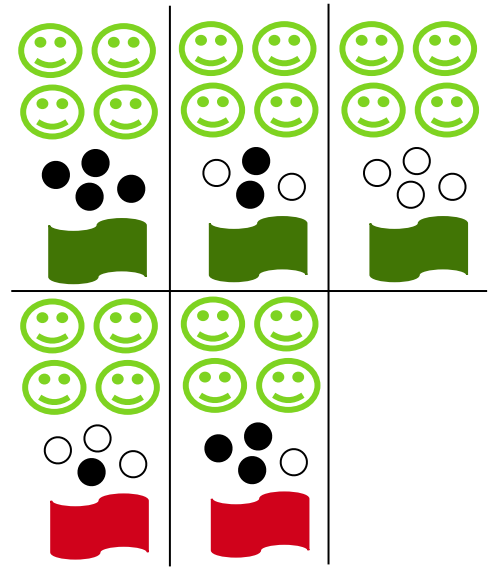}
        \caption{Even parities for the size four parity checks groups with green cards, odd parities with red ones.}
    \label{fig:evenParities}
\end{figure}

\subsection*{Inclusivity and accessibility}
In order to accommodate for a wide range of players, here are a few non-exhaustive guidelines to adapt the game to your specific audience. It must be emphasised that each player being unique, modifications must be tailored to their specific needs.

\subsubsection*{Mobility adaptation}
Different mobilities can easily be accommodated in this game, either by adapting the roles given to children with limited mobility (the needs for high-performing motor skills is less apparent in the Receiver role than it is for the Messengers of Noise ones for instance). Other options include giving time bonus/penalties to Noise versus Messengers in their crossing of the Channel.

\subsubsection*{Stimuli sensitivity adaptation}
Bearing in mind that some children might be more sensitive to loud noises, abrupt movements or physical contact, the game can easily be adapted to accommodate a wide range of sensitivities. For loud noises, children can be asked to speak words instead of shouting them, potentially making the game happen in discrete sequences. Abrupt movements could be averted by setting the rule that the Messengers and Noise can only walk across the channel.

\subsubsection*{Sensory perception and processing adaptation}
Accommodating the game for differences in sensory perception can easily be achieved with mindful planning. Examples of accommodation could include more visual communication for children who might be hearing impaired. For children with vision impairment, relying more on verbalisation can be considered, different marble textures could also be considered as an alternative to colours. For players who might be colour-blind, the choice of the two marble colours should be made considering the accessibility of the colour spectrum. 
\section*{Focus points}
This section proposes a short non-exhaustive list of items which could be paid attention to when playing. Focusing player's attention on the following question should help build an intuitive sense of the stakes of error correction in communication channels and the interplay between information redundancy and noise levels.

\begin{itemize}
    \item Why should the Noise throw a dice to decide whether they swap marble colours and not just automatically swap them? Try to allow the Receiver to observe whether the marbles are exchanged by the Noise or not. Does that change anything? If so, in which situations? Which team gets an advantage from this (if any)? Do they gain an advantage irrelevant of the Noise Level, or does the Noise Level have an impact on whether or not they do gain advantage?
    \item Does it make a difference when the ratio of Messengers versus Noise changes? What kind of difference? Does it become easier/harder for some team? Which one? Why?
    \item Does it make a difference when the Noise Value changes? What kind of difference? Does it become easier/harder for some team? Which one? Why?
    \item When \textbf{varying proportions of Messengers versus Noise}, players should be encouraged to observe how much easier/harder things get for each team. 
    \item For the \textbf{adversarial noise levels} only: interesting things to observe here is whether higher Noise Values are always good, whether keeping the Noise Values secret is an advantage for the Noise and whether changing Noise values instead of sticking to a fixed value makes it easier or harder.
    \item For the \textit{Quantum codes} version where the Receiver is left to choose the \textbf{stabiliser size}, does it become harder or easier to figure out where the error-s is/are with bigger or smaller parity check groups? Does it have anything to do with the Noise Value?
\end{itemize}

{\small
\bibliographystyle{abbrv}
\bibliography{main}

\begin{thebibliography}{10}

\bibitem{article}
A.~Bagiati, S.~Y. Yoon, D.~Evangelou, and I.~Ngambeki.
\newblock Engineering curricula in early education: Describing the landscape of open resources.
\newblock {\em Early Childhood Research and Practice}, 12, 09 2010.

\bibitem{bluvstein2024logical}
D.~Bluvstein, S.~J. Evered, A.~A. Geim, S.~H. Li, H.~Zhou, T.~Manovitz, S.~Ebadi, M.~Cain, M.~Kalinowski, D.~Hangleiter, et~al.
\newblock Logical quantum processor based on reconfigurable atom arrays.
\newblock {\em Nature}, 626(7997):58--65, 2024.

\bibitem{cheung2016quantum}
V.~Cheung and J.~Wallace.
\newblock Quantum cats: The demo.
\newblock In {\em Proceedings of the 2016 ACM International Conference on Interactive Surfaces and Spaces}, pages 445--448, 2016.

\bibitem{chiofalo2022games}
M.~L. Chiofalo, C.~Foti, M.~Michelini, L.~Santi, and A.~Stefanel.
\newblock Games for teaching/learning quantum mechanics: a pilot study with high-school students.
\newblock {\em Education Sciences}, 12(7):446, 2022.

\bibitem{chiofalo2023quantum}
M.~L. Chiofalo, J.~Yago~Malo, and L.~Gentini.
\newblock The quantum bit woman: Promoting cultural heritage with quantum games.
\newblock In {\em New Challenges and Opportunities in Physics Education}, pages 361--380. Springer, 2023.

\bibitem{dejarnette2012america}
N.~DeJarnette.
\newblock America's children: Providing early exposure to stem (science, technology, engineering and math) initiatives.
\newblock {\em Education}, 133(1):77--84, 2012.

\bibitem{Dennis_2002}
E.~Dennis, A.~Kitaev, A.~Landahl, and J.~Preskill.
\newblock Topological quantum memory.
\newblock {\em Journal of Mathematical Physics}, 43(9):4452–4505, Sept. 2002.

\bibitem{franklin2020exploring}
D.~Franklin, J.~Palmer, W.~Jang, E.~M. Lehman, J.~Marckwordt, R.~H. Landsberg, A.~Muller, and D.~Harlow.
\newblock Exploring quantum reversibility with young learners.
\newblock In {\em Proceedings of the 2020 ACM conference on international computing education research}, pages 147--157, 2020.

\bibitem{gallager1962low}
R.~Gallager.
\newblock Low-density parity-check codes.
\newblock {\em IRE Transactions on information theory}, 8(1):21--28, 1962.

\bibitem{greinert2023towards}
F.~Greinert, R.~M{\"u}ller, S.~Goorney, J.~Sherson, and M.~S. Ubben.
\newblock Towards a quantum ready workforce: the updated european competence framework for quantum technologies.
\newblock {\em Frontiers in Quantum Science and Technology}, 2:1225733, 2023.

\bibitem{greinert2024european}
F.~Greinert, R.~M{\"u}ller, S.~R. Goorney, R.~Laurenza, J.~Sherson, and M.~S. Ubben.
\newblock European competence framework for quantum technologies.
\newblock In {\em Zenodo}, pages 1--30. 2024.

\bibitem{kaur2022defining}
M.~Kaur and A.~Venegas-Gomez.
\newblock Defining the quantum workforce landscape: a review of global quantum education initiatives.
\newblock {\em Optical Engineering}, 61(8):081806--081806, 2022.

\bibitem{kitaev2003fault}
A.~Y. Kitaev.
\newblock Fault-tolerant quantum computation by anyons.
\newblock {\em Annals of physics}, 303(1):2--30, 2003.

\bibitem{kultima2021quantum}
A.~Kultima, L.~Piispanen, and M.~Junnila.
\newblock Quantum game jam--making games with quantum physicists.
\newblock In {\em Proceedings of the 24th International Academic Mindtrek Conference}, pages 134--144, 2021.

\bibitem{lazzeroni2021teaching}
C.~Lazzeroni, S.~Malvezzi, and A.~Quadri.
\newblock Teaching science in today’s society: the case of particle physics for primary schools.
\newblock {\em Universe}, 7(6):169, 2021.

\bibitem{seskirquantum}
Z.~C. Seskir, P.~Migda{\l}, C.~Weidner, A.~Anupam, N.~Case, N.~Davis, C.~Decaroli, {\.I}.~Ercan, C.~Foti, P.~Gora, et~al.
\newblock Quantum games and interactive tools for quantum technologies outreach and education: A review and experiences from the field.

\bibitem{shor1995scheme}
P.~W. Shor.
\newblock Scheme for reducing decoherence in quantum computer memory.
\newblock {\em Physical review A}, 52(4):R2493, 1995.

\bibitem{skult2021marriage}
N.~Skult and J.~Smed.
\newblock The marriage of quantum computing and interactive storytelling.
\newblock In {\em Games and Narrative: Theory and Practice}, pages 191--206. Springer, 2021.

\bibitem{thoughtXP}
M.~Violaris.
\newblock A physics lab inside your head: Quantum thought experiments as an educational tool.
\newblock In {\em 2023 IEEE International Conference on Quantum Computing and Engineering (QCE)}, volume~03, pages 58--67, 2023.

\bibitem{wolf2023brief}
R.~A. Wolf and S.~Araiba.
\newblock A brief overview of programmed instructions for quantum software education.
\newblock In {\em 2023 IEEE International Conference on Quantum Computing and Engineering (QCE)}, volume~3, pages 78--86. IEEE, 2023.

\end{thebibliography}
}

\section*{Acknowledgements}
My sincere thanks to Laurence Bouillon and Sylvain Machen for valuable discussions about the ruleset and to Marie Rodriguez for her proof-reading of inclusivity suggestions.

\end{document}